%% file: info-transfer.tex
\documentclass[aps,pre,floatfix,nofootinbib,twocolumn]{revtex4} 
\usepackage{graphicx} 
\usepackage{amsmath} 
\usepackage{amssymb}
\usepackage{subfigure}
\usepackage{booktabs}

\newcommand{\BEQ}{\begin{equation}}
\newcommand{\EEQ}{\end{equation}}
\newcommand{\BEA}{\begin{eqnarray}}
\newcommand{\EEA}{\end{eqnarray}}
\newcommand{\be}{\begin{eqnarray}}
\newcommand{\ee}{\end{eqnarray}}
\newcommand{\benn}{\begin{eqnarray*}}
\newcommand{\eenn}{\end{eqnarray*}}
\newcommand{\BGA}{\begin{gather}}
\newcommand{\EGA}{\end{gather}}

\newcommand{\comment}[1]{}

\begin{document}
\title{Information Transfer in Social Media}

\date{\today}
\author{Greg Ver Steeg and Aram Galstyan\footnote{\{gregv,galstyan\}@isi.edu}}
\affiliation{USC Information Sciences Institute,Marina del Rey, CA 90292}

\begin{abstract} 
Recent research has explored the increasingly important role of social media by examining
the dynamics of individual and group behavior, characterizing patterns of information diffusion, and identifying influential individuals. 
In this paper we suggest a measure of causal relationships between nodes based on the information--theoretic notion of transfer entropy, or information transfer. 
This theoretically grounded measure is based on dynamic information, captures fine--grain notions of influence, and admits a natural, predictive interpretation.
Causal networks inferred by transfer entropy can differ significantly from static friendship networks because most friendship links are not useful for predicting future dynamics.
We demonstrate through analysis of synthetic and real--world data that transfer entropy reveals meaningful hidden network structures. 
In addition to altering our notion of who is influential, transfer entropy allows us to differentiate between weak influence over large groups and strong influence over small groups.

\end{abstract}



\maketitle

\input{sec-intro}
\input{sec-results-synthetic}
\input{sec-results-twitter}
\input{sec-discussion}

\bibliography{network,bib-galstyan,bib-lerman,bib-lerman2}{}
\bibliographystyle{plain}

\end{document}

%% file: sec-intro.tex
\section{Introduction}
\label{sec:intro}
Recent years have witnessed an explosive growth of various social media sites such as online social networks, discussion forums and message boards, and inter-linked blogs. 
For researchers, social media serves as a fertile ground for examining social interactions  on an unprecedented scale~\cite{Castellano2009}.  
One important problem is the characterization and identification of {\em influentials}, which can be defined as  users who influence the behavior of large numbers of other users. 
Recent work on influence propagation has used numerous characterizations of influentials based on topological centrality measures such as Pagerank score~\cite{PageRank,Jeh03}. To characterize influence in Twitter, researchers have suggested number of followers, mentions, and retweets~\cite{Cha10measuringuser}, and Pagerank of follower network~\cite{Kwak2010}. It has been observed, however, that the purely structural measures of influence can be misleading~\cite{Ghosh10snakdd} and high popularity does not necessarily imply high influence~\cite{Romero2010,VerSteeg2011infotransfer}. More recent work has used  the size of the cascade trees~\cite{Bakshy11} and influence--passivity score~\cite{Romero2010}. One serious drawback of existing methods is that they are based on explicit causal knowledge (i.e., A responds to B), whereas for many data sets such knowledge is not available and needs to be discovered.

Here we suggest  a model--free approach to uncovering causal relationships and identifying influential users  based on their capacity to {\em predict} the behavior of other users, through the information-theoretic notion of {\em transfer entropy}, interchangeably referred to as information transfer. In a nutshell, transfer entropy between two stochastic processes characterizes the reduction of uncertainty in one process due to the knowledge of the other process; a mathematical definition is given below. Transfer entropy can be thought of as a nonlinear generalization of Granger causality~\cite{barnett}, and has been used extensively in computational neuroscience, e.g., for examining causal relationships in cortical neurons~\cite{Gourevitch2007}. In contrast to other correlation measures such as mutual information, transfer entropy is  asymmetric and allows differentiation in the direction of information flow. Furthermore, whereas most existing studies are concerned with {\em aggregate} measures of influence, the approach outlined here allows more fine--grained analysis of information diffusion by analyzing information transfer on each existing link in the network. Finally, our approach is model-free. Information--theoretic measures allow us to statistically characterize our uncertainty without making assumptions about human behavior. 

The rest of this paper is organized as follows.  We begin by describing the basic intuition and mathematics behind the information transfer, and briefly mention computational  issues   of the approach. In Section~\ref{sec:synthetic} we present results of our simulation with synthetically generated data, where we thoroughly examine how the information transfer depends on various characteristics of the data generating process. In Section
~\ref{sec:twitter}  we present our results on real-world data extracted from user activities on Twitter.
We conclude the paper by discussing results and some future work in Section~\ref{sec:discussion}. 

\section{Transfer Entropy} 
\subsection{Notation}
For each user, $X$, we record the history of activity, e.g., timing of tweets, as a sequence of times as $S_X = \{t_j: 0<t_1<t_2 \ldots  \}$.
In general, we assume each user's activity is described by some stochastic point process. 
We are limited by finite data to consider finite temporal resolution, so we introduce the binned random variable,
\be
B_X (a,b) \equiv 
\left\{ \begin{array}{rl} 
1 &\mbox{if }  \exists t_j \in S_X \cap (b,a],  \\ 
0 &\mbox{otherwise.}
   \end{array}\right.
\ee
If we observe the actions of a user for some long period of time $T$, we can define probabilities over these coarse-grained variables. Fix $\delta \in \mathcal{R}$, then
\benn
P(B_X(t,t-\delta) = X_{t} ) 
\equiv \frac1{T-\delta} \int_{\delta}^{T} dt [B_i(t,t-\delta) = X_{t}] 
\eenn
Similarly, we could define a joint probability distribution over a sequence of adjacent bins,
$$P(B_X(t,t-\delta_0)=X_t, B_X(t-\delta_0,t-\delta_0-\delta_1)=X_{t-1},\ldots),$$
with widths $\delta_0,\delta_1,\ldots,\delta_k \in \mathcal{R}$. We will omit the binning function for succinctness, $P(X_t, X_{t-1},\ldots,X_{t-k}).$
We can write this even more compactly by defining $X_t^{(t-k)} \equiv \{X_t,\ldots,X_{t-k} \}$.

The dynamics of a user may depend on users they are linked to in some unknown, arbitrary way. Therefore, for two users $X$ and $Y$, with activities recorded by $S_X,S_Y$, we define a joint probability distribution using a common set of bins denoted with widths $\delta_0,\delta_1,\ldots$ as 
$P(X_t^{(t-k)},Y_t^{(t-k)})$.
Conditional and marginal probability distributions are defined in the usual way and we use the standard definition for conditional entropy for discrete random variables $A,B$ distributed according to $P(A,B)$,
$$H(A | B) = -\sum_{A,B} P(A,B) \log P(A | B).$$

\subsection{Definition of transfer entropy}

The {\em transfer entropy} introduced in~\cite{Schreiber2000} is defined as
\BEQ\label{eq:TE}
T_{X\rightarrow Y} = H(Y_{t}|Y_{t-1}^{(t-k)}) - H(Y_{t}|Y_{t-1}^{(t-k)},X_{t-1}^{(t-l)})
\EEQ 
The first term represents our uncertainty about $Y_{t}$ given $Y$'s history only. The second term represents the smaller uncertainty when we know $X$'s history as well. Thus, transfer entropy explicitly describes the reduction of uncertainty in $Y_{t}$ due to knowledge of $X$'s recent activity. Note that information transfer  is asymmetric, as opposed to mutual information, and thus better suited for characterizing directed information transfer. For simplicity, we take $l = k$ from here on.

\subsection{Sampling problems and solutions}\label{sec:sample}

The use of information--theoretic techniques to analyze real-world point processes has been studied almost exclusively in the context of neural activity\cite{Victor}. Therefore, it is in this literature that the problems associated with estimating entropies for sparse point process data have been explored most thoroughly. The fundamental problem is that, in the absence of sufficient data, estimating entropies from probability distributions based on binned frequencies leads to systematic bias \cite{Panzeri}. Intuitively, if we have $k$ bins of history then we need $O(2^k)$ pieces of data in order to sample all possible histories. 

A variety of remedies are available and we make use of several. The most obvious solution is to restrict ourselves to situations where we have adequate data. In the subsequent analysis, we filter out users that are below a certain activity level. In practice, however, raising our activity threshold high enough to guarantee convergence of entropies would eliminate almost all users from our dataset. 

The next remedy to apply is to estimate the average magnitude of the systematic bias that results from using sparse data and subtract it from our estimate. When we calculate the entropies in Eq.~\ref{eq:TE}, we subtract out the Panzeri-Treves bias estimate\cite{PanzeriTreves}. Fig.~\ref{fig:pt} illustrates the effect of this bias correction as a function of amount of data collected.

The definition in Eq.~\ref{eq:TE} implicitly depends on bin widths specified by the $\delta_i$'s. The simplest procedure, and the one taken in the neural spike train literature, is to set all the bins to have equal width. We have a great deal of pre-existing empirical knowledge about human activity that can help us improve on this method. Many studies have shown that humans exhibit a heavy tail in the distribution of their response times to communications\cite{barabasi05}. This implies that bins accounting for recent activity should be narrower while bins accounting for older activity can be wider. We can even base these bin widths on measured response times, if such data is available. Using more informative bins means we can use fewer bins, reducing the effect of sampling problems.

A final technique to reduce bias is discussed in \cite{Victor} and uses a class of binless entropy estimators. These techniques carry their own mathematical difficulties and we will not consider them here.
With these tools in hand, we can proceed to use information transfer to analyze user activity in social media. 

%% file: sec-results-synthetic.tex
\section{Results}
\label{sec:results}

In this section we report the results of our experiments with both synthetic data and real world data from Twitter. 
The ultimate goal is to infer information transfer between agents in the network by analyzing their patterns of activity. Patterns of activity could include many things including timing, content, and medium of messages. We focus only on the timing of activity on 
Twitter (tweeting of URLs).  In principle, our analysis could be extended to include more complex information, but, as discussed, this would require either more data or better methods for dealing with sparse data. 

We test and validate our ability to infer information transfer from patterns of activity in two ways. 
First, while our information--theoretic analysis of social network data uses only timing of activity, the data includes unique identifiers allowing us to track the flow of information through the network. 
On Twitter, we track specific URLs. We can use the spread of these trackable pieces of information to confirm that the information transfer inferred solely from the timing of activity corresponds to actual exchanges of information.

For the synthetic data, we dictate that an agent's activity depends on its neighbors' activity in some fixed way. This allows us to check how well information transfer recovers the hidden dependence structure from activity patterns alone. For instance, even without knowing anything about the network structure, we find that a sufficient amount of data allows perfect reconstruction of the underlying network.

\subsection{Experiments with synthetic data}
\label{sec:synthetic}

To form a better understanding of different factors impacting information transfer, we performed extensive experiments with synthetically generated data. Ideally, we would like our synthetic data to reflect, in a tunable way, the challenges we face with real world data. These challenges include a long tail for human response times, heterogeneous response to neighbors' activity, background noise affecting node dynamics, incorrect data, and insufficient data. We explore these challenges first for a pair of nodes, and then for an entire network.

We model user activity as a coupled, non-homogeneous Poisson point process. Suppose that we have two nodes and a single link from $X \rightarrow Y$.  We can characterize $Y$'s activity in terms of a time-dependent rate. We define $S_X^t \equiv S_X \cap [0,t)$, that is, the activity for $X$ until time $t$. 
\BEA\label{eq:stochastic}
\lambda_Y(t|S_X^t) = \mu + \gamma \sum_{t_i \in S_X^t } g(t-t_i)
\EEA
The first term, $\mu$, represents a constant rate of background activity. The second term represents a time-dependent increase in the rate of activity in response to activity from a neighbor.  The strength of influence of $X$ is parametrized by $\gamma$. In practice, we will set the background rate equal to a constant and vary the relative strength $\gamma/\mu$ through the parameter $\gamma$. The time dependence of the influence is captured by the function $g$. 
We set $g(\Delta t) = \mbox{min}(1,\left(\frac{1 \mbox{ hour}}{\Delta t}\right)^3)$ to reflect the observed fact that the distribution of human response times are characterized by a long tail\cite{barabasi05}. 

Along with a causal network, Eq.~\ref{eq:stochastic} defines a generative model for point process activity. We can efficiently generate activity according to this model using the thinning method discussed in \cite{Ogata}. We vary the total amount of data by fixing the background rate $\mu = 1 \mbox{ event/day}$ and varying the total amount of observation time, $T$. Equivalently, we could have fixed $T$ and varied the rate of activity. 
After fixing the parameters, we can generate data and then use that data to infer the appropriate probabilities to calculate information transfer according to Eq.~\ref{eq:TE}. 

\begin{figure}[htbp] 
\centering
\subfigure[]{
    \includegraphics[width=2.5in]{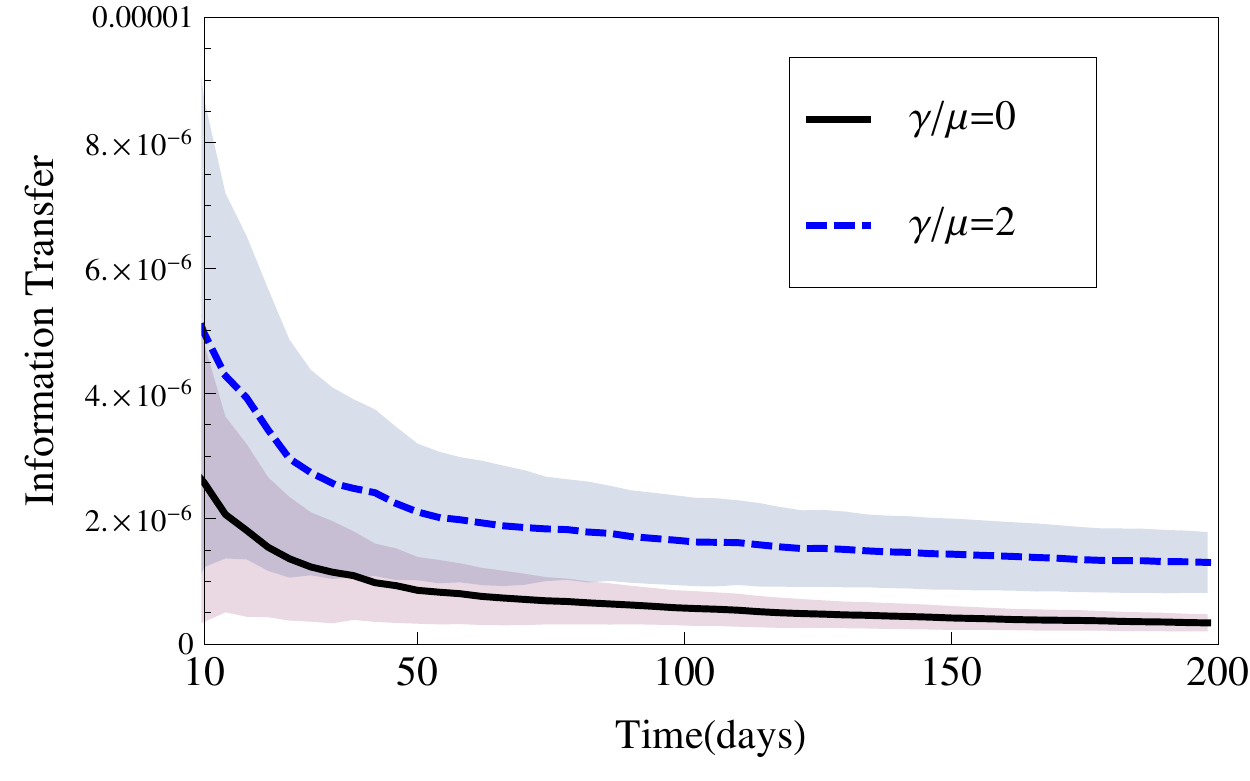} \label{bias}
    } 
    \subfigure[]{
    \includegraphics[width=2.5in]{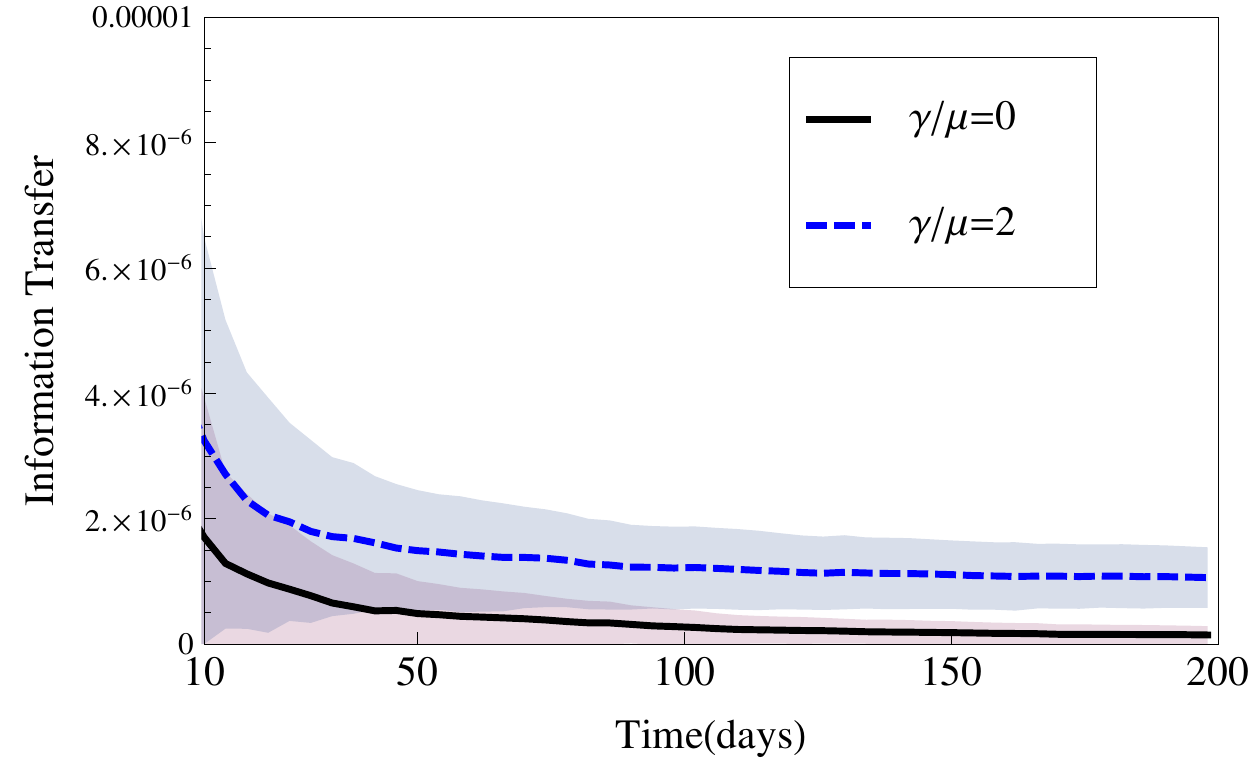} \label{nobias}
    }
   \caption{Mean and std for the estimate of information transfer averaging over 200 pairs of users with $\gamma/\mu=0,2$ as a function of time. (a) Results without correcting for bias and (b) with Panzeri-Treves bias correction\cite{PanzeriTreves}.}
   \label{fig:pt}
\end{figure}

As discussed in Sec.~\ref{sec:sample}, we take a variety of measures to ensure good estimation. In this case, we directly control the amount of data through the parameter $T$. For the bin widths we choose $\delta_0 = 1\mbox{ sec}$, fixing the finest temporal resolution. For the history we choose wider bin widths for less recent history. In the synthetic examples we take the past three hours of history into account by choosing $\delta_1 = 1\mbox{ hour},\delta_2 = 2\mbox{ hours}$. Also, it should be assumed that the Panzeri-Treves bias estimate has been taken into account, except in Fig.~\ref{bias} where we compare results without bias correction.

\begin{figure}[htbp] 
   \centering
   \includegraphics[width=2.5in]{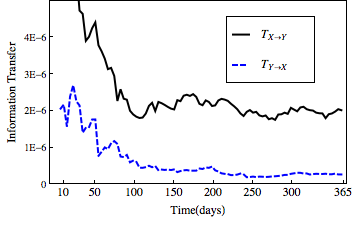} 
   \caption{If we have influence from $X \rightarrow Y$ but not vice versa, the asymmetry in the information transfer correctly reflects the direction of influence. Information transfer plotted for a single pair of users.}
   \label{fig:asymmetry}
\end{figure} 

Note that in the example in Eq.~\ref{eq:stochastic}, we have allowed $X$ to affect $Y$, but not vice versa. As a first test we can generate some data for a pair of users and then compare $T_{X \rightarrow Y}$ and $T_{Y \rightarrow X}$. In Fig.~\ref{fig:asymmetry}, we compare these two quantities when $\gamma/\mu = 2$ as a function of the total observation time $T$.

In Fig.~\ref{fig:pt} we examine the accuracy and convergence of information transfer estimates as a function of time both with and without bias correction. We ran 200 trials and plot the mean and standard deviation of the information transfer estimate at each time step. Clearly, there is a systematically high estimate in the low sampling regime, but, even in that case, higher influence leads to a higher information transfer on average. The Panzeri-Treves bias correction drastically reduces, but does not completely eliminate, this systematic error. 

%

\begin{figure}[htbp] 
   \centering
   \includegraphics[width=2.5in]{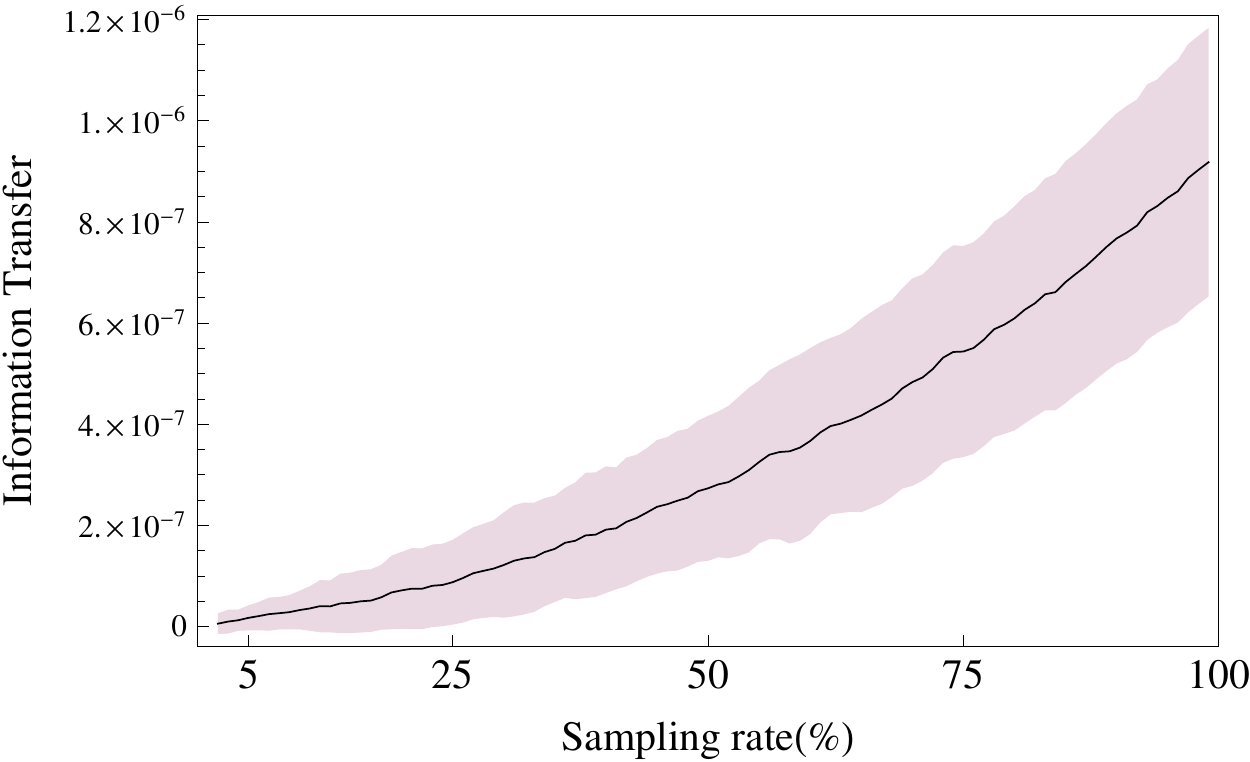} 
   \caption{A summary of the mean and std of the inferred value of $T_{Y \rightarrow X}$ averaged over 200 trials as a function of the sampling rate, with $T=500$ days and $\gamma/\mu=2$.}
   \label{fig:summary}
\end{figure}

Next, we consider the same scenario, where we generate $X,Y$ according some stochastic process, but now imagine that we do not see all activity. That is, what if we do not see every event due to limited sampling? This is often the case, for instance, with Twitter data, where researchers typically have access to only a small fraction of all tweets, ranging from $1\% - 20\%$. So we set a sampling parameter $f$, and say that for each $t_i \in S_X$, we only keep that event with probability $f$.  A summary of how the final transfer entropy, $T_{Y \rightarrow X}$, depends on the sampling rate, $f$, is given in Fig.~\ref{fig:summary}. We show the results after $500$ days to guarantee enough data to be very close to convergence. We see that sampling drastically reduces the inferred transfer entropy, destroying our ability to deduce flow of information. 

\begin{figure}[htbp] 
   \centering
   \includegraphics[width=0.8 \columnwidth]{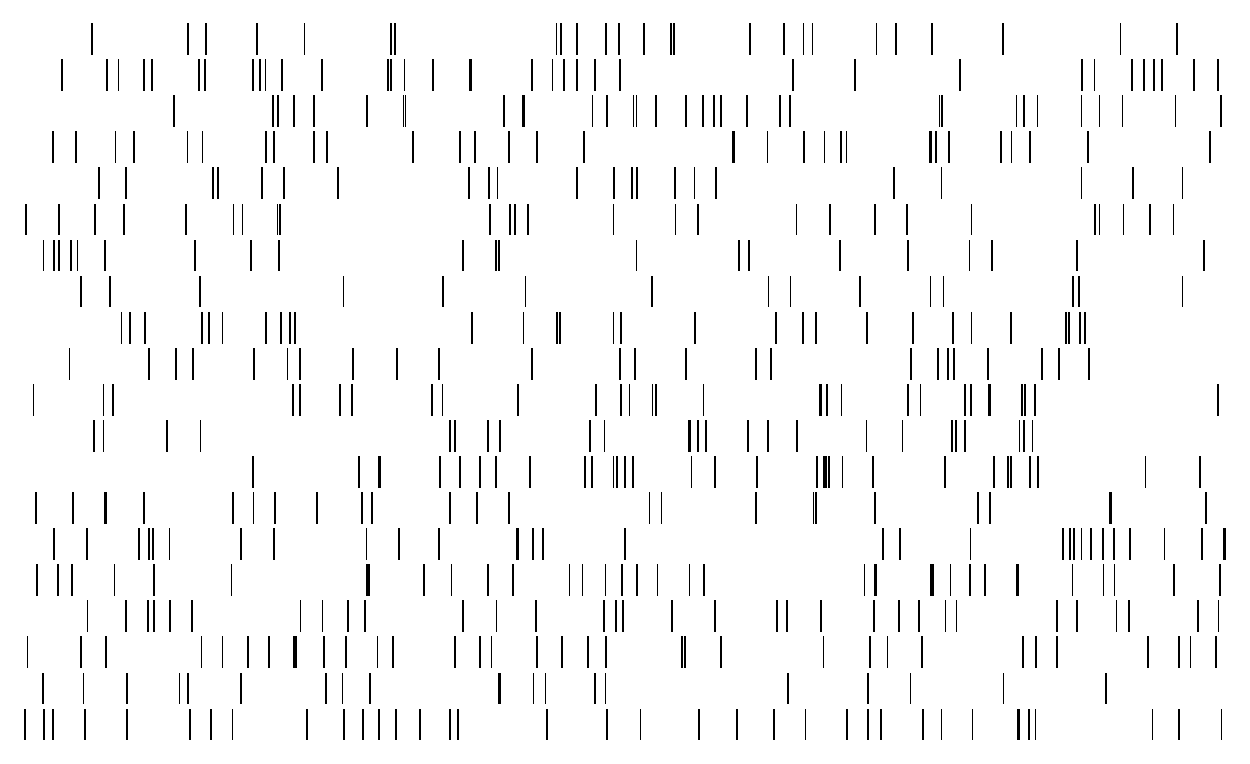} 
   \caption{Each row represents a different user. Each line represents an event for that user over a time period of thirty days. With enough data we could calculate the information transfer between each pair of users and recover the unknown network structure exactly.}
   \label{fig:activity}
\end{figure}

So far, we have only considered two nodes with a single link between them. Now, we want to consider a  directed, causal network of $N$ nodes, with some arbitrary connectivity pattern. We consider a similar stochastic model as defined in Eq.~\ref{eq:stochastic}, except now we denote the set of $Y$'s neighbors (i.e., people who can influence $Y$) as $\mathcal{N}(Y)$.
\BEA\label{eq:stochastic2}
\lambda_Y(t|S_{\mathcal{N}(Y)}^t) = \mu +  \sum_{X\in\mathcal{N}(Y)} \gamma_X \sum_{t_i \in S_X^t } g(t-t_i)
\EEA
To begin we imagine $\gamma_X=\gamma$ for all neighbors, but in general a node may be affected more strongly by some neighbors than others. A sample of activity generated according to this model is given in Fig.~\ref{fig:activity}. 

\begin{figure}[!htb]
\centering
\subfigure[]{
    \includegraphics[width=0.4\textwidth]{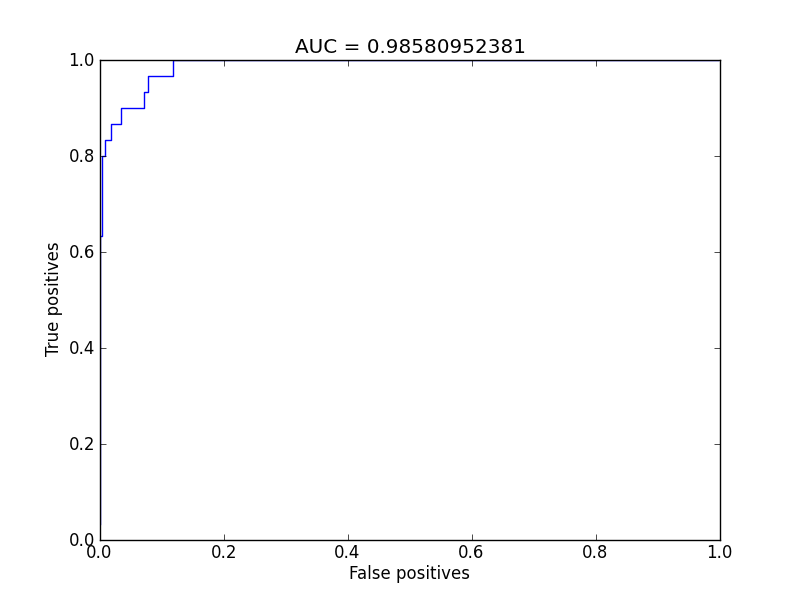} \label{fig_v}
    } 
    \subfigure[]{
    \includegraphics[width=0.4\textwidth]{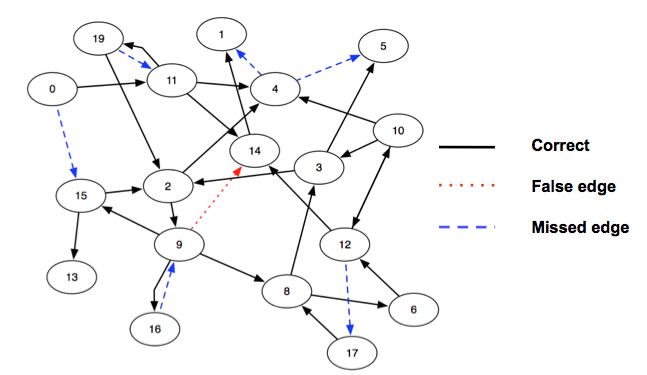} \label{fig_c}
    }
     \caption{(a) ROC curve and (b) transfer--entropy induced graph for the synthetically generated data described in the text. Threshold is chosen according to F-measure. Black solid lines correspond to true positives, red dashed lines to false positives and blue dotted lines to false negatives. }
     \label{fig_0}
\end{figure}

The challenge is to take the information given by the activity and recover the underlying graph structure. For each pair of nodes, $X,Y$, we calculate $T_{X \rightarrow Y}$. Then we pick some threshold $T_0$, and if $T_{X \rightarrow Y} > T_0$, we consider there to be an edge from $X \rightarrow Y$, otherwise not. We could check our true positive rate and false positive rate as a function of $T_0$, as shown in Fig.~\ref{fig_v}, for $N=20, \gamma/\mu = 1.0$ and time = $450$ days. We show an example of the recovered versus actual network in Fig.~\ref{fig_c}, using a threshold picked according to F-measure.

\begin{figure}[htbp] 
   \centering
   \includegraphics[width=2.5in]{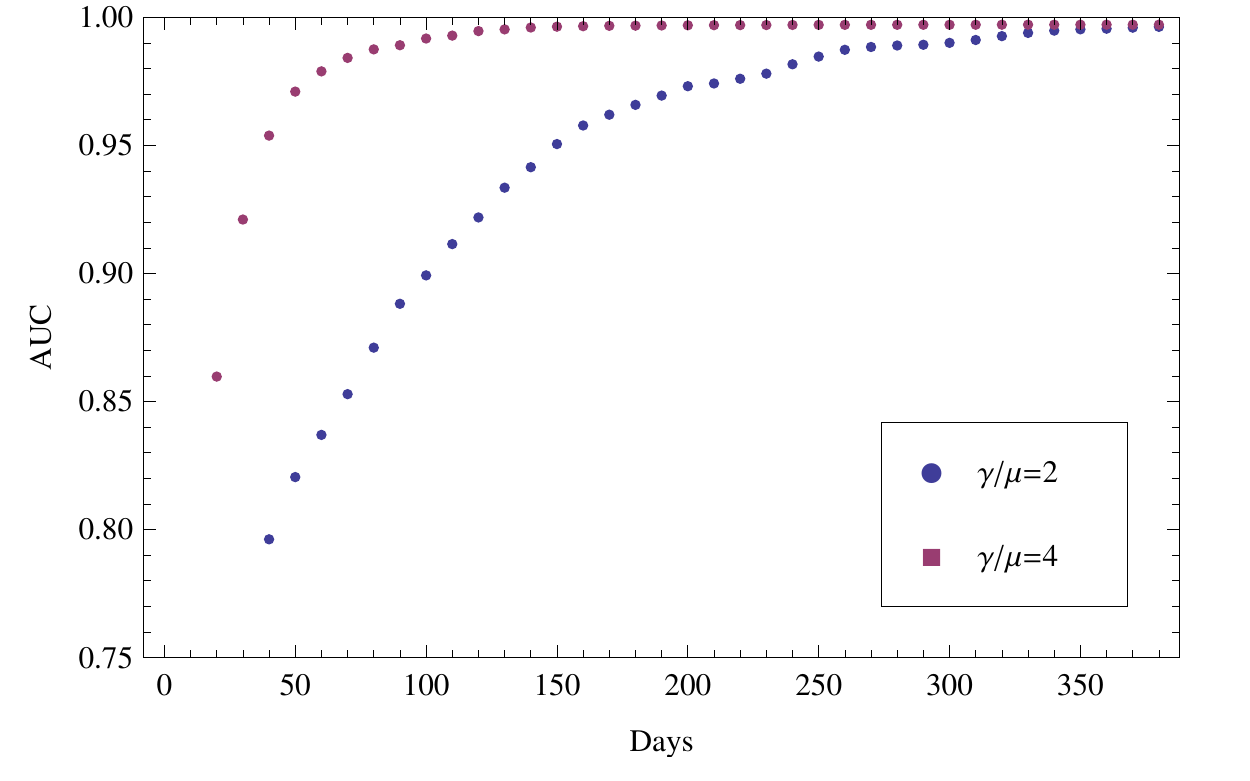} 
   \caption{AUC of the network inferred using transfer entropy as a function of $T$, with $\gamma/\mu=2,4$.}
   \label{fig:auc_summary}
\end{figure}

The previous example was chosen to show what kinds of errors arise given a weak signal. In general, with either enough data or strong enough influence, we can perfectly recover the underlying graph structure. If we consider the area under the ROC curve (AUC), as in Fig.~\ref{fig_v}, then an AUC of 1 corresponds to perfect reconstruction of the graph. We summarize the AUCs for random networks with $N=20$ and $\langle k \rangle =3$, while varying $T$ and $\gamma/\mu$ in Fig.~\ref{fig:auc_summary}.

\begin{figure}[htbp] 
   \centering
   \includegraphics[width=2.5in]{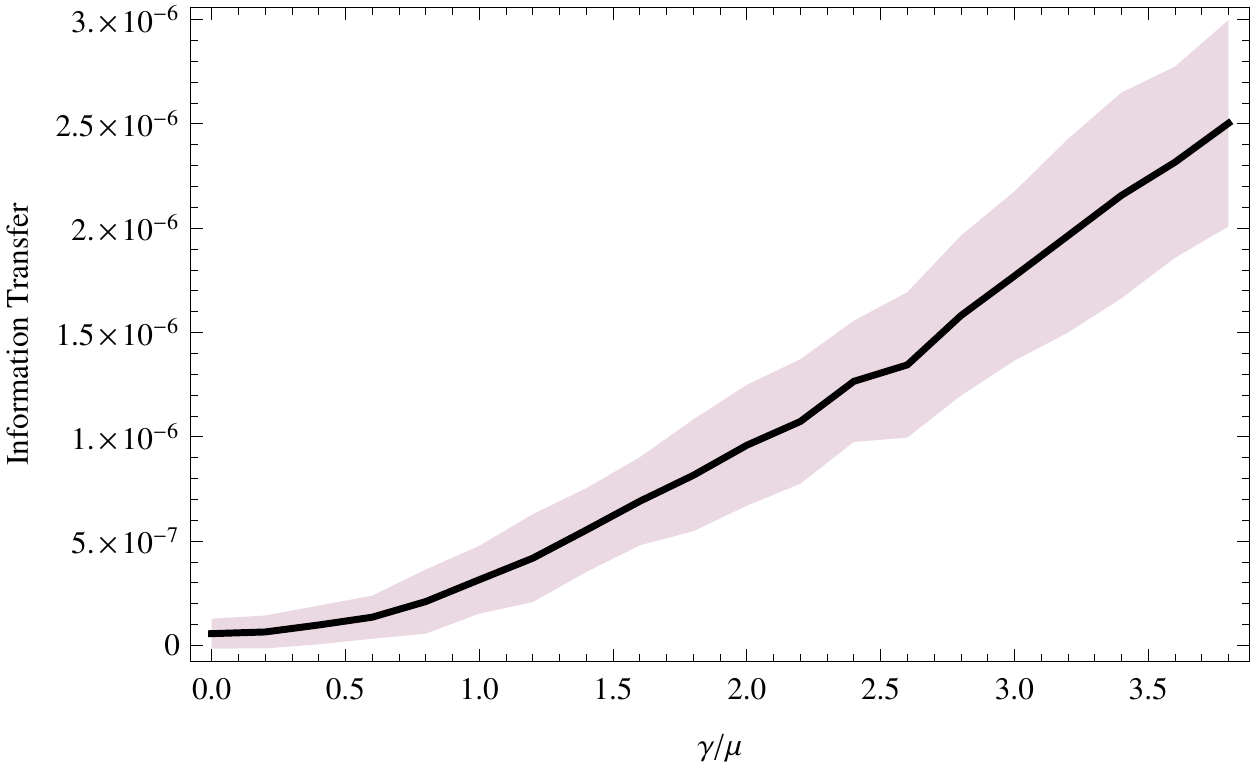} 
   \caption{Information transfer between pairs of nodes for varying $\gamma/\mu$ with $T=500\mbox{ days}$. The black line corresponds to the mean information transfer for a given $\gamma/\mu$ and the shaded region denotes the standard deviation after 100 trials. }
   \label{fig:weights}
\end{figure}

As a final experiment, we can consider the effect of allowing different $\gamma$ between different pairs of nodes. Fig.~\ref{fig:weights} shows that transfer entropy is able to recover the relative influence well.

In principle, there are many other effects we could have considered to make a more realistic synthetic model. Background and influence rates should vary for different individuals. There may be periodicity defined by daily, weekly, and monthly cycles. However, because information transfer makes no model assumptions, it is relatively insensitive to such details. The main constraint is data, which is why we focused on sensitivity to amount and quality of observations. 

%% file: sec-results-twitter.tex
\subsection{Results for Twitter dataset}
\label{sec:twitter}
Twitter is a popular micro-blogging service. As of July 2011, users  send 200 million tweets per day. Twitter has become an important tool for researchers both due to the volume of activity and because of the easily available tools for data collection. Twitter's ``Gardenhose'' API, allows access to $20\%-30\%$ of all tweets. 

Unfortunately, 
as discussed in Sec.~\ref{sec:synthetic}, 
filtering of data can lead to a drastic reduction in the measured information transfer. Instead, the Gardenhose API was used to identify URLs being tweeted. Then, the search API was used to find all mentions of these URLs in any tweets by any users. In this way, the filtering limitation is avoided, while we restrict ourselves to the domain of URL posting. Additionally, each URL corresponds to a unique piece of information whose movement through the network can be traced. The data also includes the full social network among ``active users'', in this case, anyone who tweeted a URL in the three week collection period. The data we used was collected in the fall of 2010 \cite{Ghosh10snakdd}. The dataset included about 70 thousand distinct URLs, 3.5 million tweets, and 800 thousand users. We further filtered our results to ``very active'' users, namely, users who tweeted at least 10 URLs during this time period. 

Before we can calculate transfer entropy as presented in Eq.~\ref{eq:TE}, we need to specify the relevant bin widths. We take the finest resolution to be $\delta_0 = 1\mbox{ second}$, the same resolution as presented by the Twitter API. For binning of the history, we used distribution of observed re-tweet response times to motivate a choice of $\delta_1 = 10\mbox{ min}, \delta_2 = 2\mbox{ hours}, \delta_3 = 24\mbox{ hours}$. Although we saw a long tail of re-tweet times stretching into days, our data were insufficient to include this weak effect. By limiting ourselves to only three bins, we only have to sample over 8 possible histories. 
Note that the activity is for any tweeting of URLs; our calculations do not make use of the information encoded in the URL.
We then calculate the transfer entropy between each pair of users who are connected. 

\begin{figure}[htbp] 
   \centering
   \includegraphics[width=2.5in]{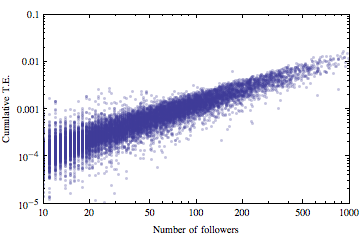} 
   \caption{For each user, we compare the number of their followers to their cumulative outgoing transfer entropy. Note that the outgoing transfer entropy may differ by an order of magnitude for people with the same number of followers.}
   \label{fig:out_v_te}
\end{figure}

The result of this procedure is the construction of a directed, weighted graph, where each edge in the original directed graph is now labeled by the calculated transfer entropy. We can now compare standard measures of influence to measures based on this weighted graph.
The simplest measure of influence on static graphs is to count the number of followers a user has. This ignores the fact that not all followers are the same, nor do followers react in the same way to different people that they follow. For instance, it may be that a recommendation from a close friend is worth more to a person than the same recommendation from five acquaintances. This problem is only exacerbated by the recent emergence of ``followers for pay'' services, which seek to artificially inflate the number of followers to your Twitter account. In Fig.~\ref{fig:out_v_te}, we explore the comparison between out degree and transfer entropy and we find that although on average people with more followers have more transfer entropy, two people with the same number of followers may have vastly different influence as measured by transfer entropy. 

\begin{figure}[htbp] 
   \centering
   \includegraphics[width=2.5in]{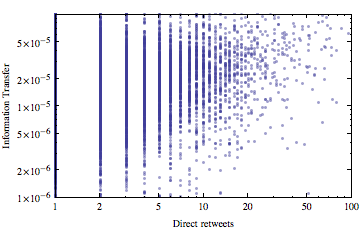} 
   \caption{The number of URLs that were first tweeted by user X and subsequently tweeted by X's follower, Y, is correlated with the calculated transfer entropy between X and Y, even though transfer entropy is calculated only from the timing of activity, without regard for specific URLs. Pearson's correlation coefficient is 0.22. }
   \label{fig:twitter_validation}
\end{figure}

To verify that transfer entropy is a meaningful quantity, we could test how well the transfer entropy, based only on the timing of activity, matches the measured flow of information, as determined by tracing specific URLs. To that end, for each pair of connected users, $X \rightarrow Y$, we count how many specific URLs were first tweeted by $X$ and then subsequently re-tweeted by $Y$. This number is compared to the transfer entropy in Fig.~\ref{fig:twitter_validation}. The existence of even a weak correlation is surprising considering the limited amount of data and the fact the transfer entropy is not making use of URL or re-tweet information at all. 
We also note that while a high number of re-tweets implies high information transfer, a low number of re-tweets is uncorrelated with information transfer. This makes sense because information transfer measures influence that is not necessarily in the form of re-tweets; we will give some examples below.

Table~\ref{fig:topit} shows the edges with the highest information transfer. These accounts are all solely for the purpose of promotion. Taking the top example, for instance, reveals that these two accounts will tweet exactly the same message within a few seconds of each other. Note that in the text of their tweets neither account uses re-tweets or an ``@'' for attribution. Twitter specifically forbids indiscriminate automatic re-tweets and has a policy against duplicate accounts. Many of the accounts on this list have since been banned by Twitter. 
\begin{table}[htbp]
   \centering
   \begin{tabular}{@{} lcr @{}} 
      \toprule
      \cmidrule(r){1-2} 
     {\bf User}    & {\bf Follower} & {\bf I.T.($\cdot 10^{-6}$)}\\
      \midrule
Free2BurnMusic&Free2Burn&4328\\
Earn\_Cash\_Today&income\_ideas&1159\\
BuzTweet\_com&scate&1006\\
Free2Burn&Free2BurnMusic&939\\
Kamagra\_drug2&sogradrug3&929\\
sougolinkjp&sogolinksite&903\\
kcal\_bot&FF\_kcal\_bot&902\\
nr1topforex&nr1forexmoney&795\\
wpthemeworld&wpthememarket&709\\
viagrakusurida&viagrakusuride&679\\
BoogieFonzareli&Nyce\_Hunnies&668\\
A\_tango&kobuntango&662\\
Kamagra\_drug2&sogra\_drug3&638\\
dti\_affiliate&kekkonjyoho&630\\
Best\_of\_Deals&Orbilook\_SMI&621\\
viagrakusurida&kamagra\_100mg3&561\\
kcal\_bot&Family\_Mart&542\\
kamagra\_100mg3&viagrakusuride&535\\
viagra\_drug&baiagura\_drug&532\\
kcal\_bot&Seven\_Eleven\_&530\\
      \bottomrule
   \end{tabular}
   \caption{List of edges with highest information transfer. All are promotion accounts and many of the accounts have been banned since the data were collected.}
   \label{fig:topit}
\end{table}

\begin{figure}[!htb]
\centering
\subfigure[]{
    \includegraphics[width=0.4\textwidth]{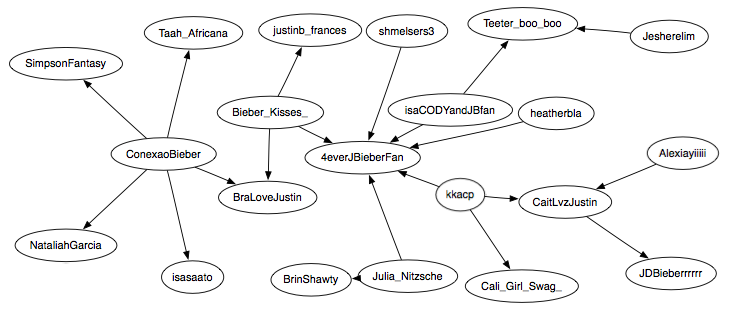} \label{bieber}
    } 
    \subfigure[]{
    \includegraphics[width=0.4\textwidth]{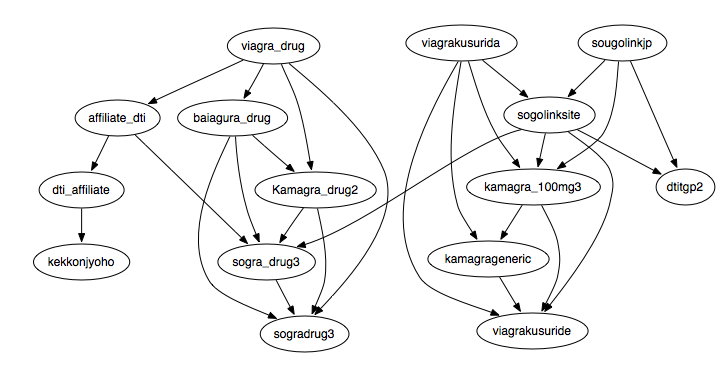} \label{drugs}
    }
     \subfigure[]{
    \includegraphics[width=0.3\textwidth]{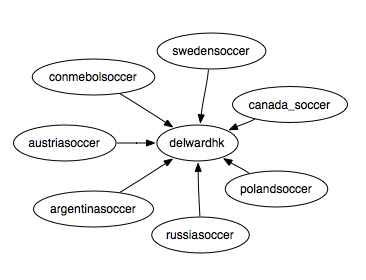} \label{soccer}
    }
     \caption{(a) This cluster appears to be non-automated, and revolves around fandom of singer Justin Bieber. (b) The cluster of drug spam accounts. (c) An account which aggregates soccer news by following and re-tweeting different regional soccer accounts. }
     \label{fig:twittergraphs}
\end{figure}

To see more complex examples, we restrict ourselves to the top 1000 edges according to information transfer. Then we look at the largest connected components. The largest component involved 600 users in Brazil, most of whom had multiple tweets of the form ``BOMBE O SEU TWITTER, COM MILHARES DE NOVOS FOLLOWERS, ATRAVES DO SITE: http://? \#QueroSeguidores'', where ``?'' was a frequently changing URL.  Google translates this as ``Pump up your Twitter, get thousands of new followers, link to this site: http://? \#IWantFollowers.'' Clicking on some of these links suggests that this a ``followback'' service. You agree to follow previous users who have signed up and in return other users of the service follow your account.  It also appears from the text that you are required to re-tweet the link to get your followers. Some other examples of high information transfer clusters are shown in Fig.~\ref{fig:twittergraphs}.

\begin{figure}[htbp] 
   \centering
   \includegraphics[width=2.5in]{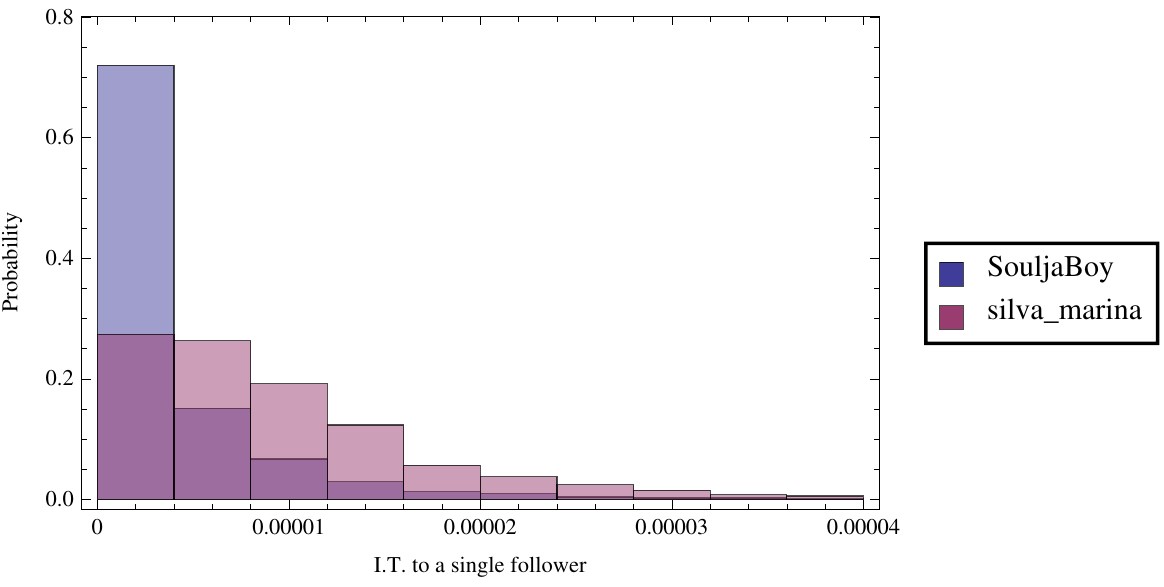} 
   \caption{A histogram showing the probability distribution of outgoing transfer entropy for followers of two different Twitter accounts.}\label{fig:silva}
\end{figure}

We consider another advantage of measuring influence through information transfer by looking at two users who had almost the same outgoing transfer entropy ($\sim 0.025$, in the top 20 for individuals in our dataset), but vastly different behavior of followers. The first Twitter account is SouljaBoy, a prominent American rapper who is also very active in social media. 
The second account is ``silva\_marina'', the Twitter account of Marina Silva, a popular Brazilian politician.  This data was taken during the run up to the Brazilian presidential election, in which Marina Silva was a candidate; she received $19.4\%$ of the popular vote. At first it seems surprising that the SouljaBoy, who has six times the followers, should have a similar outgoing transfer entropy to a politician known mostly in one country and with fewer than a million Twitter followers. On the other hand, Fig.~\ref{fig:silva} reveals the reason for this disparity. Marina Silva may have fewer followers, but her effect on them tended to be much stronger. Marina Silva's activity tended to be a better predictor of her followers' behavior than Soulja Boy's activity was for his followers.

The strength of Marina Silva's influence along with the serendipitous timing before the Brazilian elections suggests another intriguing possibility. It seems likely that not only does transfer entropy vary for different followers, it may vary over time as well. This suggests that a dynamic estimate of information transfer could detect changes in the importance of individuals in the network.

%% file: sec-discussion.tex
\section{Discussion}
\label{sec:discussion}

We have presented a novel  information--theoretic approach for measuring influence. In contrast to previous studied that focused on aggregate measures of influence, the transfer entropy used here allows us to characterize and quantify the causal information flow for any pair of users.   For a small number of users, this can allow us to reconstruct the network of connections from user activity alone. For large networks, this allows us to identify the most important links in the network. 

The method used here for calculating information transfer did not require any explicit causal knowledge in the form of re-tweets or other textual information. On the one hand, this may be an advantage in situations where such information is either missing or misleading, as was the case in the example for marketers on Twitter. On the other hand, we may be neglecting valuable information, and in the future we would like to incorporate textual information in more sophisticated ways but still within an information--theoretic approach. 
Although this should be straightforward in principle, in practice entropy based approaches require large amounts of data. More complex signals require a commensurate increase in data. Therefore, the other main thrust of future work should be towards reducing data required for entropy estimation, either through better bias correction or through binless approaches\cite{Victor}.

Because this measure has a rigorous interpretation in terms of predictability, it allows us to easily understand results that might otherwise seem anomalous. For instance, in one example we found that Marina Silva, the Brazilian presidential candidate, had high information transfer both to and from a Brazilian news service. Neither Twitter account ever retweeted or explicitly mentioned a tweet of the other. However, there was an external cause, the upcoming debates and elections, that explains both of their activities. Without knowing this external cause, it is entirely consistent to say that either user's activity could help you predict the others. In fact, it may be possible to use this bi-directional predictability to identify external causes in the first place.

Another result that is easy to understand in the context of predictability is the high incidence of ``spam'' in our results. This is no surprise since a large amount of spam is produced by automated systems and these systems are intrinsically very predictable. Although identifying spam is a natural application of our analysis, some human behavior stood out as well. Diehard fandom also leads to quite predictable behavior. 

Many existing notions of influence are static, ill-defined, ad hoc, or only apply in aggregate. Information transfer is a rigorously defined, dynamic measure capable of capturing fine-grain notions of influence and admitting a straightforward predictive interpretation. Many of the mathematical techniques necessary have already been developed in the neuroscience literature and we have shown how to usefully adapt them to a social media context. 

\subsection*{Acknowledgments}
We would like to thank Armen Allahverdyan for useful discussions. This research was supported in part by the National Science Foundation under grant No. 0916534 and US AFOSR MURI grant No. FA9550-10-1-0569.